\documentclass{article}
\usepackage{ijcai18}

\usepackage{times}
\usepackage{xcolor}
\usepackage{soul}
\usepackage[utf8]{inputenc}
\usepackage[small]{caption}

\usepackage{helvet}  %Required
\usepackage{courier}  %Required
\usepackage{url}  %Required
\usepackage{graphicx}  %Required
\usepackage{amsmath}
\usepackage{amssymb}
\usepackage{multicol}  
\usepackage{multirow} 
\usepackage{algorithm} 
\usepackage{algorithmic} 
\usepackage{graphicx}
\usepackage{latexsym} 

\title{Ensemble Neural Relation Extraction with Adaptive Boosting}

\author{Jérôme Lang\\ 
Laboratoire d'Analyse et Modélisation des Systèmes pour l'Aide à la Décision (LAMSADE)  \\
pcchair@ijcai-18.org}

\author{
Dongdong Yang$^1$, 
Senzhang Wang$^{2}$\thanks{Corresponding author} 
and
Zhoujun Li$^{3}$
\\ 
$^1$ Unversity of Southern California \\
$^2$ Nanjing University of Aeronautics and Astronautics\\
$^3$ Beihang University  \\
dongdony@usc.edu,
szwang@nuaa.edu.cn,
lizj@buaa.edu.cn
}

\begin{document}

\maketitle

\begin{abstract}
Relation extraction has been widely studied to extract new relational facts from open corpus. Previous relation extraction methods are faced with the problem of wrong labels and noisy data, which substantially decrease the performance of the model. In this paper, we propose an ensemble neural network model - Adaptive Boosting LSTMs with Attention, to more effectively perform relation extraction. Specifically, our model first employs the recursive neural network LSTMs to embed each sentence. Then we import attention into LSTMs by considering that the words in a sentence do not contribute equally to the semantic meaning of the sentence. Next via adaptive boosting, we build strategically several such neural classifiers. By ensembling multiple such LSTM classifiers with adaptive boosting, we build a more effective and robust joint ensemble neural networks based relation extractor. Experiment results on real dataset demonstrate the superior performance of the proposed model, improving F1-score by about 8\% compared to the state-of-the-art models.
\end{abstract}

\section{Introduction}
Many NLP tasks have been built on different knowledge bases, such as Freebase and DBPedia. However, the knowledge bases could not cover all the facts in the real world. Therefore, it is essential to extract more common relational facts automatically in open domain corpus. As known, relation extraction (RE) aims at extracting new relation instances that are not contained in the knowledge bases from the unstructured open corpus. It aligns the entities in the open corpus with those in the knowledge bases and retrieves the entity relations from the real world. For example, if we aim to retrieve a relation from the raw text, ``\emph{Barack Obama married Michelle Obama 10 years ago}'', a naive approach would be to search the news articles for indicative phrases, such as ``\emph{marry}'' or ``\emph{spouse}''. However, the result may be wrong since human language is inherently various and ambiguous. 

Previous supervised RE methods require a large amount of labelled relation training data by human-hand. To address this issue, Mintz et al. \cite{mintz2009distant} proposed an approach via aligning the entity in KB for later extraction without plenty of training corpus. However, their assumption - there is only one relation existing in a pair of entities, was irrational. Therefore, later researches assumed more than one relation could exist between a pair of entities. Hoffmann et al. \cite{hoffmann2011knowledge} proposed a multi-instance learning model with overlapping relations (MultiR) that combined a sentence-level extraction model for aggregating the individual facts. Surdeanu et al. \cite{surdeanu2012multi} proposed a multi-instance multi-label learning model (MIML-RE) to jointly model the instances of a pair of entities in text and all their labels. The major limitation of the above methods is that they cannot deeply capture the latent semantic information from the raw text. It is also challenging for them to seamlessly integrate semantic learning with feature selection to more accurately perform RE.

Recently, deep neural networks are widely explored for relation extraction and have achieved significant performance improvement \cite{zeng2015distant,linneural}. Compared with traditional shallow models, deep models can deeply capture the semantic information of a sentence. Zeng et al. \cite{linneural} employed CNN with sentence-level attention over multiple instances to encode the semantics of sentences. Miwa and Bansal \cite{miwa2016end} used a syntax-tree-based long short-term memory networks (LSTMs) on the sentence sequences. Ye et al.\cite{ye2017jointly} proposed a unified relation extraction model that combined CNN with a pair of ranking class ties. However, the main issue of existing deep models is that their performance may not be stable and could not effectively handle the quite imbalanced, noisy, and wrong labeled data in relation extraction even if a large number of parameters in the model. %For this reason, we could regard neural networks under this case as weak learners.

\begin{figure*}[htbp]
\centering
\includegraphics[width=16.9cm]{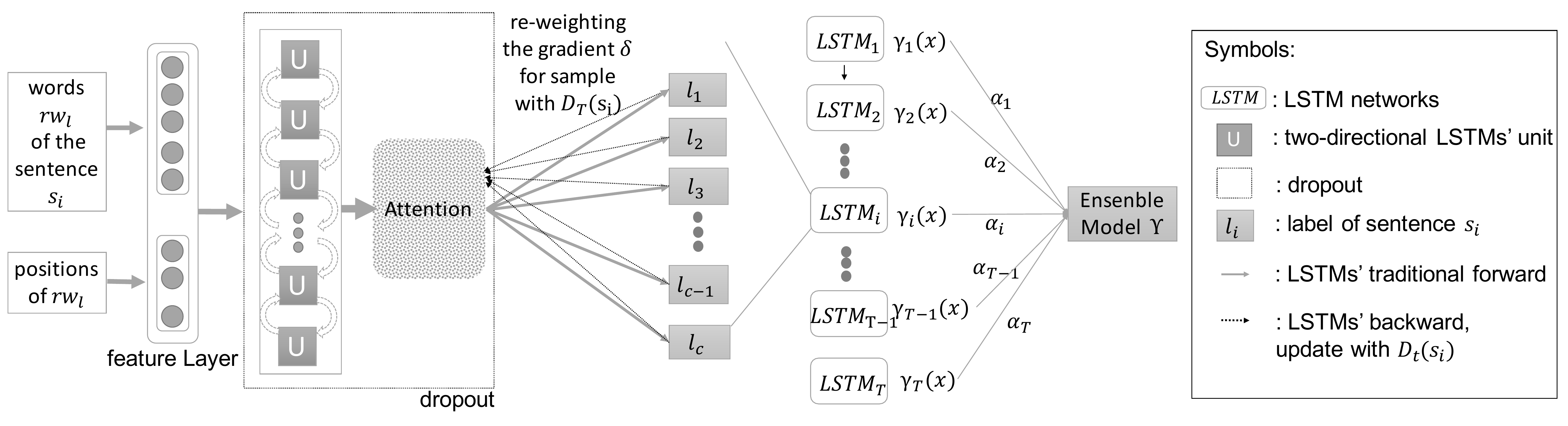}
\caption{The framework of Ada-LSTMs contains three layers: feature layer, bi-directional Stacked LSTMs' layer with attention and adaptive boosting layer. $s_i$ indicates the original input sentence with a pair of entities and their relation.}
\label{fig:new-arch}{}
\end{figure*}

To address the above issues, in this paper we propose a novel ensemble deep neural network model to extract relations from the corpus via an Adaptive Boosting LSTMs with Attention model (Ada-LSTMs). Specifically, we first choose bi-directional long short-term memory networks to embed forward and backward directions of a sentence for better understanding the sentence semantics. Considering the fact that the words in a sentence do not contribute equally to the sentence representation, we import attention mechanism to the bi-directional LSTMs. Next we construct multiple such LSTM classifiers and ensemble their results as the final prediction result. Kim and Kang \cite{kim2010ensemble} showed that ensemble with neural networks perform better than one single neural network in prediction tasks. Motivated by their work, we import adaptive boosting and tightly couple it with deep neural networks to more effectively and robustly solve the relation extraction problem. The key role of adaptive boosting in our model is re-weighting during the training process. The weight of incorrectly classified samples will increase. In other words, the samples classified wrongly gain more attention so that the classifier is forced to focus on these hard examples. Note that attention can distinguish the different importance of words in the sentence, while adaptive boosting can use sample weights to inform the training of neural networks. In a word, the combination of the two can more precisely capture the semantic meaning of the sentences and better represent them, and thus help us train a more accurate and robust model.

We summarize the contributions of this paper as follows.
\begin{itemize}  

\item We propose a Multi-class Adaptive Boosting Neural Networks model, which to our knowledge is the first work that combines adaptive boosting and neural networks for relation extraction.

\item We utilize adaptive boosting to tune the gradient descent in NN training. In this way, a large number of parameters in a single NN can be learned more robustly. The ensembled results on multiple NN models can achieve more accurate and robust relation extraction result.

\item We evaluate the proposed model on a real data set. The results demonstrate the superior performance of the proposed model which improves F1-score by about 8\% compared to state-of-the-art models.

\end{itemize}

\section{Related Work}

As an important and fundamental task in NLP, relation extraction has been studied extensively. Many approaches for RE have been developed including distant supervision, deep learning, etc.  Distant supervision was firstly proposed to address this issue by \cite{mintz2009distant}. Mintz et al. aligned Freebase relations with Wikipedia corpus to automatically extract instances from a large-scale corpus without hand-labeled annotation. Riedel et al. \cite{riedel2010modeling} tagged all the sentences with at least one relation instead of only one. Hoffmann et al. \cite{hoffmann2011knowledge} also improved the previous work and aimed at solving the overlapping relation problem. Surdeanu et al. \cite{surdeanu2012multi} proposed a multi-instance multi-label method for relation extraction. Zheng et al. \cite{zheng2016aggregating} aggregated inter-Sentence information to enhance relation extraction.

With neural networks bursting out many fields of research, researchers also began to apply this new technique to relation extraction. Zeng et al. \cite{zeng2014relation} first proposed a convolutional neural network (CNN) for relation classification. Zhang et al. \cite{zhang2015bidirectional} proposed to utilize bidirectional long short-term memory networks to model the sentence with sequential information about all words. 

Recently, attention has been widely used in NLP tasks. Yang et al. \cite{yang2016hierarchical} used a two-layer attention mechanism for document classification, which inspires us to focus on the word understanding level. Liu et al \cite{linneural} also used the attention level to its CNN architecture and gained a better performance in extraction. \cite{zhang2017position} combines an LSTM sequence model with a form of entity position-aware attention for relation extraction. 

Besides, ensemble learning is a well-known machine learning paradigm which tries to learn one hypothesis from training data, ensemble methods. Freund et al. \cite{freund1996experiments} was the first paper which proposed Adaboost. Rokach et al. \cite{rokach2010ensemble} showed us the technique of generating multiple models strategically and combining these models to improve the performance of many machine learning tasks. Li et al. \cite{li2017cross} also showed that ensemble technique can be successfully used in transfer learning.

\section{Methodology}

Given a sentence $s_i \in \mathcal S$, where $\mathcal S$ is a corpus, and its corresponding pair of entities $\varphi=(e_1, e_2)$, our model aims at measuring the probability of each candidate relation $\Omega_i \in \mathcal L$. $\mathcal L$ is defined as \{$1, 2, 3, ..., C$\}, where $C$ is the number of relation classes.

Figure \ref{fig:new-arch} shows the overview of the model framework. The model mainly consists of three layers: feature layer, bi-directional LSTMs layer with an attention and adaptive boosting layer. The feature layer makes sentence vectorized and embeds them as the input of the model. The bi-directional LSTMs with attention layer can deeply capture the latent information of each sentence. Attention mechanism could weight each phase in a sentence, which is learned during the training process. The adaptive boosting layer combines multiple classifiers to generate the final weighted joint function for classifying the relation. The essential notations used in this paper and their meanings are given in Table \ref{tab:tab0}. Next, we will introduce the three layers of the proposed model in details in the following sections.

\begin{table}\small
\begin{center}  
\begin{tabular}{c|p{6cm}}
\hline  
Notations    & Interpretation\\
\hline
$Y$ 		& the final trained classification model\\ 
$\gamma_t(x)$ 	& a neural network classifier\\  
$T$ 		& the number of trained neural network classifiers \\  
$\alpha_t$ 	& the weight of the neural classifier $\gamma_t(x)$\\  
$D_t$ 		& the weight vector for total samples at the $t^{th}$ epoch\\
$D_t(s_i)$ 	& the weight for sentence $s_i$ at the $t^{th}$ epoch\\
$s_i$		& the $i^{th}$ sentence in the corpus $\mathcal S$\\ 
$rw_k$		& the $k^{th}$ word in a sentence $s_i$\\
%$x_k$		& the word embedding for the $k^{th}$ word\\
%$d^p_k$		& the position embeddings for the $k^{th}$ word\\
\hline  
\end{tabular}
\caption{Notations and their meanings.}  

\label{tab:tab0}
\end{center}
\end{table}

\subsection{Embedded Features}
The embedded features contain word embeddings and position embeddings. We use two embedded features for relation extraction as the input of the bi-directional long short-term memory neural networks. We describe the embedding features as follows.

\subsubsection{Word Embeddings}
The inputs are some raw words \{$rw_1$,$rw_2$,...,$rw_l$\}, where $l$ is the length of the input sentence. We make every raw word $rw_i$ represented by a real-valued vector $w_i$ via word embedding which is encoded by an embedding matrix $M$ $\in$ $\mathbb{R}^{d^a \times V}$, where $V$ is the representation of a fix-sized vocabulary and $d^a$ is the dimension of the word embedding. In our paper, we use the skip-gram model %\cite{mikolov2013distributed} 
to train word embeddings.

\subsubsection{Position Embeddings}
A position embedding is defined as a word distance, which is from the position of the word to the positions of the entities in a sentence. A position embedding matrix is denoted as $P \in {\mathbb{R}^{l^p \times d^p}} $, where $l^p$ is the number of distances and $d^p$ is the dimension of the position embedding proposed by Ye et al. \cite{ye2017jointly}. As there are two entities in a sentence that we need to measure their distances to the word, we have two $d^p$ values. Therefore, the dimension of the word representations is $d^w=d^a+2 \times d^p$ and the final input vector for raw word $rw_i$ is $x_i=[w_i,d^p_1,d^p_2]$.

\iffalse
\subsubsection{Part-of-speech tags}
Due to the information may not agree well with the specific sentence, we introduce POS into embedded-sentence features through aligning each word with its POS tag, e.g., noun, verb, etc.

\subsubsection{Grammatical relations}
Since we are using dependency tree to extract the most important information at the same time ignore unimportant part in a sentence, the relations between a governing word and its children ought to be considered for the neural networks could attain better understanding of a sentence. For example, Salisbury $\in$ England, instead of England $\in$ Salisbury.
\fi

\subsection{Multi-class Adaptive Boosting Neural Networks}

The Multi-class Adaptive Boosting Long Short-term Memory Neural Networks (Ada-LSTMs) is a joint model, in which several neural networks are combined together according to their weight vector $\alpha$, learned from adaptive boosting algorithm. Before describing the model in detail, we would like to show the motivation for coming up with this model. We first analyze the distribution of a public dataset for relation extraction, which is currently widely used as the benchmark and released by \cite{riedel2010modeling}. The data distribution is quite unbalanced. Among the 56 relation ties, 32 of them have less than 100 samples and 12 of them have more than 1000 samples. Besides, as \cite{liu2016learning} discussed, the dataset has wrong labelling and noisy data problems. Thus it is difficult for a single model to achieve promising result on relation extraction with such noisy and distorted training data. Therefore, it is essential to introduce a robust algorithm to alleviate the wrong labelling data issue and the distortions of the data.

In our model, we adopt multi-class adaptive boosting method to improve the robustness of the neural networks for relation extraction. For the neural networks part, we use LSTMs because it is naturally suitable to handle the sequential words in a sentence and captures the meanings well. For the ensemble learning part, Adaboost is a widely used ensemble learning method that sequentially trains and ensembles multiple classifiers. The $t^{th}$ classifier is trained with more emphasis on different weights on the input samples, which is based on a probability distribution $D_t$ to re-weighing the samples. The original adaptive boosting \cite{freund1996experiments} is to solve the binary classification problems and calculate the samples one by one. To make it fit into our model, we make the following modifications as shown in Equations (\ref{eq:ada1})-(\ref{eq:ada9}).

\begin{equation}
\label{eq:ada1}
Y(x)=f(\sum^T_t{\alpha_t \gamma_t(x)})
\end{equation}

The final prediction model $Y$ is obtained by weighted voting as shown in (\ref{eq:ada1}), where $\alpha_t$ means the weight of each classifier $\gamma_t(x)$ for our final extractor $Y(x)$. The softmax function $f$ in Equation (\ref{eq:ada1}) is to predict the labels of relation types. Here we focus more on the upper level of the model architecture, and more details about the neural classifier $\gamma_t(x)$ will be given in the next section. The result of training the $t^{th}$ classifier is such a hypothesis $h_t:X \rightarrow L$ where $X$ is the space of input features and $L=\{1, ..., c\}$ is the space of labels. 

\begin{equation}
\label{eq:ada3}
\alpha_t=\frac{1}{2}ln\frac{1-\epsilon_t}{\epsilon_t}
\end{equation}

The weight $\alpha_t$ of each NN classifier $\gamma_t(x)$ is updated based on its training error $\epsilon$ on the training set as shown in Equation (\ref{eq:ada3}). After the $t^{th}$ round the weighted error $\epsilon_t$ of the resulting classifier is calculated.

During the training process, the weight $\alpha$ of each classifier is learned by a parameter vector $D_t$, which is the sentence weight for total samples in one epoch. Different from \cite{freund1996experiments} assigned equal value for each sentence in the dataset, our model assigns equal value for each batch in the dataset. Each batch contains the same number of sentences. The vector $D_1(b_i)=\frac{1}{n}$, where $n$ is the number of batches, $b_i$ is the $i^{th}$ batch of all the samples. $D_1$ means the initialized vector parameter in the first epoch. In our case we process the samples batch by batch. 

\begin{equation}
\label{eq:ada4}
\epsilon=\sum_j{e^{\tau_j}}, \tau < \frac{1}{2}
\end{equation}

\begin{equation}
\label{eq:ada5}
\tau_j = \frac{\sum^K_kerror(I(\gamma(k)\neq y_k))}{K}
\end{equation}

The Equations (\ref{eq:ada4})-(\ref{eq:ada5}) show how to calculate the training error $\epsilon$, which is used for updating the vector paramter $D_t$. $error(I(\gamma(k)\neq y_k))$ means that when the model output $\gamma(k)$ is not equal to its true label $y_k$ in a batch $b_i$, we gather the error of that batch. $I$ is the error indicator and $K$ is the batch size. Finally we average the error $\tau$ of each batch $j$ as shown in Equation (\ref{eq:ada5}). 

\begin{equation}  
\label{eq:ada6} 
c(x) = \left\{ \begin{array}{rl}
	e^{\alpha_t}, & \tau < \frac{1}{2}	\\
	e^{-\alpha_t}, & \tau \geq \frac{1}{2} 
	\end{array}
	\right.
\end{equation}

\begin{equation} 
D_{t+1}(b_i)=\frac{D_t(b_i)}{Z_t}c(x)
\label{eq:ada7} 
\end{equation}

After calculating the weight $\alpha$ of each classifier, we could use it to update the vector $D_t$ as shown in Equations (\ref{eq:ada6})-(\ref{eq:ada7}), where $Z_t$ is a normalization constant. $D_t$ is the weight vector for the samples at the epoch $t$. $D_{t+1}$ is computed from $D_t$ by increasing the probability of incorrectly labeling samples. We maintain the weight $D_t(b_i)$ for the batch $b_i$ during the learning process. Then, we could use it to inform the training process of neural networks, by setting a constraint to gradient descent during back propagation of neural networks. By combing Equations (\ref{eq:ada3})-(\ref{eq:ada7}), we have Equation (\ref{eq:ada8}).

\begin{equation}
\label{eq:ada8}
D_{t+1}(b_i)=\frac{D_t(b_i)}{Z_t}e^{-\alpha_t y_i \gamma_t(x_i)}
\end{equation}

More details about how re-weighting affects the neural networks are given in the following. During the training process, if the training samples are trained enough and has been fitted well, its weight $D_t(b_i)$ will drop. Otherwise, the weight $D_t(b_i)$ will increase if the samples are classified wrong so that it could contribute more to the gradient descent during training the model. That means, on each round the weight of incorrectly classified samples are increased so that the classifier is forced to focus on the hard examples \cite{freund1996experiments}. The weights $D_t$ affect the the gradient descent in back propagation during training. We assign the parameter $D_t(b_i)$ to the gradient of the back propagation as shown in Equation (\ref{eq:ada9}). Then the neural networks' parameters are updated via back propagation with $D_t$, as the architecture Figure (\ref{fig:new-arch}) shows. In this way, the adaptive boosting algorithm informs the neural networks. We learn $D_t$ as the weights of the samples to impact the neural networks. Finally, multiple NN classifiers are learned and combined as a joint relation extractor.

\begin{equation}
\label{eq:ada9}
\delta_{new} = \delta_{old} \times D_t \times \beta
\end{equation}

\begin{algorithm}[!ht]  
    \caption{Ada-LSTMs Model for Relation Extraction.}
	\label{alg:alg1} 

	\hspace*{0.02in} {\bf Input:} 
	($s_1$, $\varphi_1$, $\Omega_1$), ($s_2$, $\varphi_2$, $\Omega_2$),...,($s_m$, $\varphi_m$, $\Omega_m$), where $s_i \in \mathcal S$ is a sentence in the sentence set $\mathcal S$, $\varphi_i$ is a pair of entities and $\Omega_i \in \mathcal L$ is their relation tie.

	\hspace*{0.02in} {\bf Output:} 
	final weighted extrator $Y(x)$

    \begin{algorithmic}[1]

    	\FOR{$t=1$ to $T$}  

     	\STATE init $D_t$ on $\{1,...,n\}$
    	
    	\FOR{$\mathrm s$ in $\mathcal S$}
    
    	\STATE look up embedding $x$ for words in $s$

    	\STATE $Att\text{-}LSTMs$ FORWARD ($x$)
    	\STATE update $\delta$ based on Equation (\ref{eq:ada9})
    	\STATE $Att\text{-}LSTMs$ BACKWARD

    	\STATE calculate training error $\epsilon_t$ of $\gamma_t$:
    	\STATE \quad $\epsilon_t = Pr_{D_t}[\gamma_t(x_i) \neq y_i]$
        \STATE select classifier with smallest error $\epsilon_t$ on $D_t$

    	\STATE calculate $\alpha_t$, $c(x)$ based on Equation (\ref{eq:ada3})-(\ref{eq:ada6})
    	\STATE $D_{t+1}=g(D_t, \alpha_t, \Omega, \gamma_t)$
   		\STATE $\gamma_t:X \rightarrow {\mathcal L}$
		\ENDFOR 
        \ENDFOR  
        \STATE final prediction model: $Y(x)=f(\sum^T_t{\alpha_t \gamma_t(x)})$
    \end{algorithmic}  
\end{algorithm}

The pseudocode of the Ada-LSTMs model %\footnote{The code of our model is publicly available at https://github.com/RE-2018/re} 
is given in Algorithm \ref{alg:alg1}. $m$ is the total number of training data. $n$ is the number of batches. $\epsilon_t$ is the training error of the training samples. $\delta_{old}$ is the final layer derivative in back propagation and $\delta_{new}$ is the new derivative in back propagation used to update the networks. $\beta=\frac{1}{max(D_t)}$ is the reciprocal of maximal $D_t(b_i)$ value, where $i$ is the index of batch number and $s$ is a sentence in the corpus $\mathcal S$. $\beta$ is a coefficient, aiming at avoiding $D_t$ too small to update the NN. $g$ is a mapping function. $f$ is a softmax function. $Att\text{-}LSTMs$ is the LSTMs with selective attention model, which will be described later.

\subsubsection{LSTMs with Selective Attention (Att-LSTMs)}

In this part, as shown in algorithm \ref{alg:alg1}, we elaborate more details of the proposed neural networks with selective attention (Att-LSTMs), which more specifically is attention-based long short-term neural networks. The recursive neural networks have shown in marvelous priority in modeling sequential data \cite{miwa2016end}. Therefore, we make use of LSTMs to deeply learn the semantic meaning of a sentence which is composed of a sequence of words for relation extraction.

\begin{equation}  
\label{eq:lstm1} 
{
\left (  
             \begin{array}{c}  
             i_t\\  
             f_t\\
             o_t\\
             g_t
             \end{array}  
\right ) 
 = 
\left (  
             \begin{array}{c}  
             \sigma \\
             \sigma \\
             \sigma \\
             tanh
             \end{array}  
\right ) T_{d^w+d, d}
\left (  
             \begin{array}{c}  
             x_t \\
             h_{t-1} 
             \end{array}  
\right )
}
\end{equation}  

\begin{equation} 
\begin{array}{c}
\displaystyle c_t = f_t \odot c_{t-1} + i_t \odot g_t \\
\displaystyle h_t = o_t \odot tanh(c_t)
\end{array}
\label{eq:lstm2} 
\end{equation}

The LSTM's unit is summarized in Equation (\ref{eq:lstm1}). A sentence is initially vectorized into a sequence of encoded words $\{x_1, x_2, ..., x_l\} \in \mathbb{R}^{d^w}$, where $l$ and $d^w$ are the lengths of the input sentence and the dimension of word representations, respectively. $d$ represents the LSTM dimensionality. As Equations (\ref{eq:lstm1})-(\ref{eq:lstm2}) show, $i_t$, $f_t$, $c_t$, $o_t$, $h_t$ are the input, forget, memory, output gate and hidden state of the cell at time $t$, respectively. The current memory cell state $c_t$ is the combination of $c_{t-1}$ and $g_t$, weighted by $i_t$ and $f_t$, respectively. $\sigma$ denotes a non-linear activation function. $\odot$ means the elements-wise multiplication. $d$ denotes the dimensionality of LSTM. In our implementation of relation extraction, an input sentence is tagged with the target entities and the relation type. For further usage, we concatenate the current memory cell hidden state vector $h_t$ of LSTM from two directions as the output vector $h_k$=[$\overrightarrow{h_t}$,$\overleftarrow{h_t}$] at time $t$. Combining two directions of the sentence could better utilize the features to predict the relation type.

\begin{equation} 
 \alpha_{t}=\frac{exp(e_t)}{\sum^l_{t=1}exp(e_t)}=\frac{exp(f_{a}(h_t))}{\sum^l_{t=1}exp(f_{a}(h_t))}
\label{eq:lstm4} 
\end{equation}

\begin{equation} 
c=\sum^l_{t=1}{\alpha_{t} h_t}
\label{eq:lstm5} 
\end{equation}

We add an attention model \cite{xu2015show} to neural networks. The idea of attention is to select the most important piece of information. Since not all words contribute equally to the sentence representation, the important meaning of the sentence could be presented by the informative words to form a more effective vector representation via attention. Finally, we dropout our architecture on both attention layer and bi-directional LSTMs layer. The attention mechanism is shown in Equations (\ref{eq:lstm4})-(\ref{eq:lstm5}). For each word location $t$, $f_{a}$ is a function learned during training. Specifically, $e_t=f_{a}(h_t)=\sigma(Wh_t+b)$, where $W$ and $b$ will be learned during training and $\sigma$ is a non-linear function. Then we get a normalized importance weight $\alpha_t$ through a softmax function. $l$ means the length of the sentence sample. Then, we compute the sentence vector $c$ as a sum of adaptive weighted average of state sequence $h_t$. In this way, we could selectively integrate information word by word with attention.

\begin{equation} 
L_0=-\sum_{k=1}^n\sum_{i=1}^C{y_{ki}}log(q_{ki})
\label{eq:lstm6} 
\end{equation}

Finally, we use cross entropy \cite{de2005tutorial} to design our loss function $L_0$ as shown in Equation (\ref{eq:lstm6}). $n$ is the total number of samples. $C$ is the number of labels. $q=f(c)$, where $c$ is the output of attention layer and $f$ is the softmax function. Our training goal is to minimize $L_0$.

\subsection{Implementation Details}

\subsubsection{Learning Rate}
We followed the method referred to \cite{kingma2014adam} to decay the learning rate. The adaptive learning rate decay method is defined as $lr_t \Leftarrow lr_{t-1} *  \frac{\sqrt{1-\beta_2^2}}{1-\beta_1^t}$, where $lr_t$, $lr_{t-1}$ are the current and the last learning rates, respectively.

\subsubsection{L2 Regulation}
L2 regulation imposes a penalty on the loss goal $L_0$. For the training goal, we use a negative log likelihood of the relation labels for the pair of entities as function loss. The L2 regulation is as $L2={\lambda}\sum^n_{i=1}{W_i^2}$. It should have the same order of magnitude so that L2 regulation would not weight too much or too little in the training process. We set the constant $\lambda$ based on the above rule.

\section{Experiments}

\subsection{Dataset}
We evaluate our model on the public dataset \footnote{http://iesl.cs.umass.edu/riedel/ecml/}, which is developed by \cite{riedel2010modeling}. The dataset was generated via aligning the relations in Freebase with the New York Times corpus (NYT). The dataset induces the relationship for entities of NYT corpus into 56 relationships. The training part is gained by aligning the sentences from 2005 to 2006 in NYT and contains 176,662 non-repeated sentences, among which there are 156,662 positive samples and 20,000 no-answer (NA) negative samples. The testing part is gained in 2007 and contains 6,944 non-repeated samples, among which there are 6,444 positive samples and 500 NA negative samples.

\begin{table}
\begin{center}  

%\begin{tabular}{{p{6cm}|c}}  
\begin{tabular}{{c|c}}  
\hline  
Number of epochs & 40  \\  
LSTMs' unit size & 350 \\  
Dropout probability & 0.5 \\  
Batch size & 50 \\  
Position dimension & 5 \\
Word dimension & 50 \\
Unrolled steps of LSTMs & 70\\  
Number of neural networks & 20 \\
Initial learning Rate & $10^{-3}$ \\
L2 regulation Coefficient & $10^{-4}$ \\
\hline  
\end{tabular}  
\caption{Parameter Settings}  
\label{tab:tab1}
\end{center}  
\end{table}

\begin{table*}[tp] \small
	\centering
	
    \begin{tabular}{|c|c|c|c|c|c|c|c|c|c|c|c|c|}  
    \hline  
    \multirow{2}{*}{P@N(\%)}&  
    \multicolumn{4}{c|}{One}&
    \multicolumn{4}{c|}{Two}&
    \multicolumn{4}{c|}{All}
    \cr\cline{2-13}  
    &100&200&300&Avg&100&200&300&Avg&100&200&300&Avg
    \cr  
    \hline  
    \hline  
    CNN+ATT 	&76.2		&65.2		&60.8		&67.4		&76.2		&65.7		&62.1	&68.0	&76.2	&68.6	&59.8 	&68.2\cr\hline  
   	PCNN+ATT 	&73.3		&69.2		&60.8		&67.8		&77.2		&71.6		&66.1	&71.6	&76.6	&73.1	&67.4	&72.2\cr\hline  
   	Rank+ExATT 	&-			&-			&-			&-			&-			&-			&-		&-		&83.5 	&82.2 	&78.7	&81.5\cr\hline
    Ada+LSTM 	&\bf82.0	&\bf81.0	&\bf76.7	&\bf79.9	&\bf85.0	&\bf80.5	&\bf77.6&\bf81.0&\bf95.0 	&\bf92.5  	&\bf92.0	&\bf93.1\cr  
    \hline  
    \end{tabular}  
    \caption{P@N comparison with state-of-the-art methods.}  
    \label{tab:performance_comparison}
\end{table*}

\subsection{Experiment Settings}

\subsubsection{Word Representations}
Similar to \cite{ye2017jointly}, we keep the words that appear more than 100 times to construct word dictionary. In our paper, the vocabulary size of our dataset is 114,042. We use word2vec\footnote{http://code.google.com/p/word2vec} to train the word embedding on the NYT corpus. We set word-embedding to be 50-dimensional vectors. Additionally, the vectors will concatenate two position embedding, 2 $\times$ 5 dimensional vector, as its final word embedding.

\subsubsection{Hyper-parameter settings}

Table \ref{tab:tab1} shows the parameter settings. We set some parameters empirically, such as the batch size, the word dimension, the number of epochs. We set the weights of L2 penalty as $10^{-4}$ and the learning rate as $10^{-3}$, which both are chosen from $\{10^{-1}, 10^{-2}, 10^{-3}, 10^{-4}, 10^{-5}\}$. We select 350 LSTM's units based on our empirically parameter study from the set $\{250, 300, 350, 400, 450\}$. The selection for the number of classifiers will be discussed in the experiment results.

\subsection{Evaluation}

To evaluate the proposed method, we select the following state-of-the-art feature-based methods for comparison through held-out evaluation:

\textbf{\emph{Mintz}} \cite{mintz2009distant} is a traditional distant supervised model via aligning relation data on Freebase.

\begin{figure}[htbp]
\centering
\includegraphics[width=7.5cm]{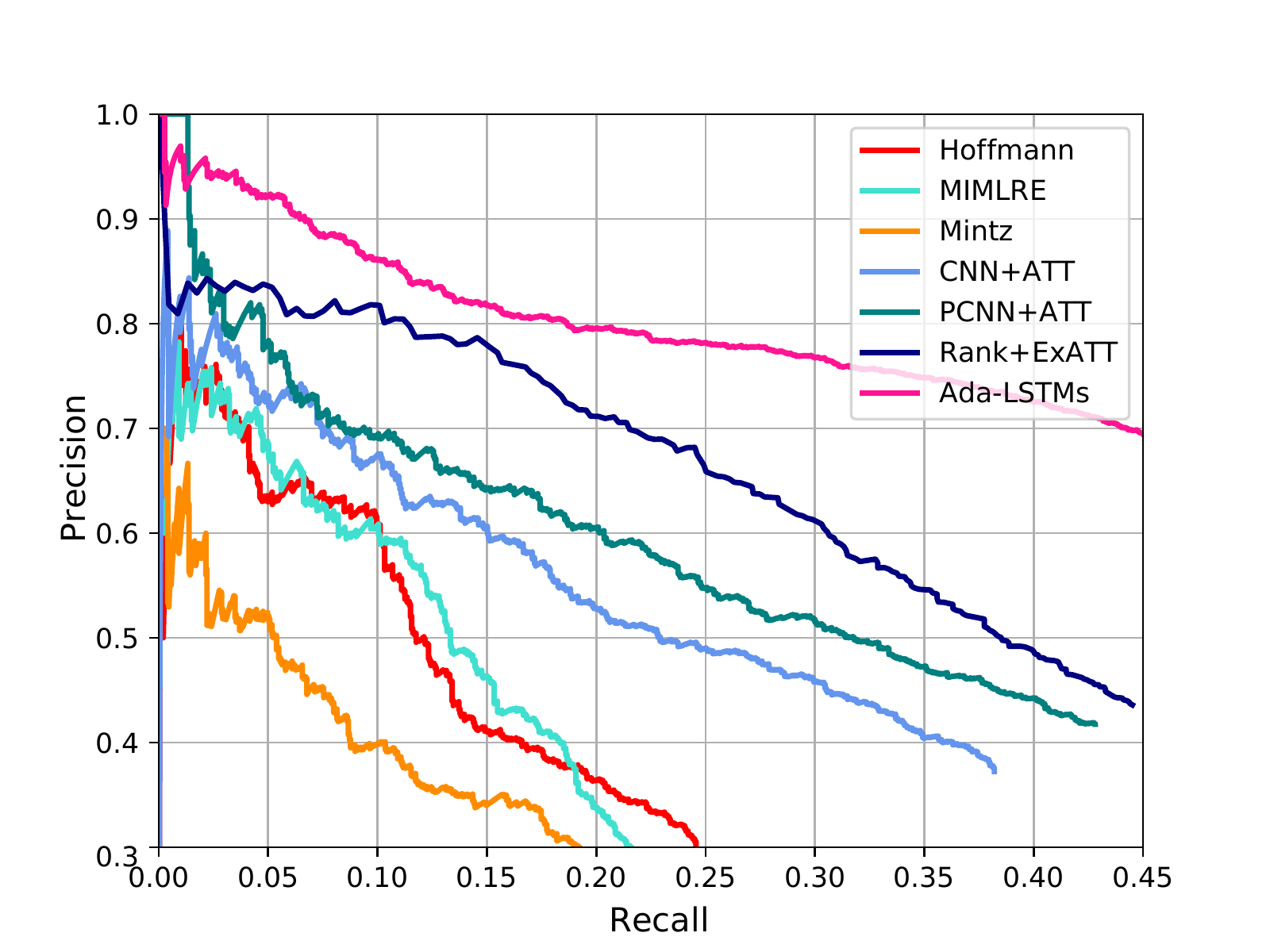}
\caption{Precision-Recall curves of different methods.}
\label{fig:result_adaboost}
\end{figure}

\textbf{\emph{MultiR}} \cite{hoffmann2011knowledge} is a graphical model of multi-instance to handle the overlapping relations problem.

\textbf{\emph{MIML}} \cite{surdeanu2012multi} jointly models both multiple instances and multiple relations.

\textbf{\emph{CNN+ATT, PCNN+ATT}} \cite{linneural} add attention mechanism to CNN and PCNN models which are proposed by Zeng et al. \cite{zeng2014relation,zeng2015distant}. Compared to CNN, PCNN adopts convolutional architecture with piecewise max pooling to learn relevant features.

\textbf{\emph{Rank+ExATT}} \cite{ye2017jointly} aggregates ranking method to attention CNN model.

In our experiments, we run the model nearly 40 epochs. Each epoch has 3533 steps (batches). At the first 10 epochs, the loss of the model drops quickly and then the loss becomes relatively stable. Therefore, in the following experiments, we select the Ada-LSTMs model with 10-30 rounds training steps as the final joint extractor for relation extraction.

We compare our Ada-LSTMs model with the above baselines and the Precision Recall (PR) curves are shown in Figure \ref{fig:result_adaboost}. From the result, one could conclude that:

(1) Our proposed method Ada-LSTMs outperforms all the baseline methods. The F1-score of our model is 0.54, which is the highest and outperforms the latest state-of-the-art model Rank+ExATT by nearly $\bf{8\%}$.

(2) Our method Ada-LSTMs has a more robust performance because the precision-recall curve is more smooth than other methods. With the increase of recall, the decay tendency of precision is obviously slower than others. Especially when recall is low, the precision of Ada-LSTMs still performs well unlike the others dropping rapidly.

We next evaluate our model via the precision@N(P@N), which means the top N precisions of the results, as shown in Table \ref{tab:performance_comparison}. One, Two, All mean that we randomly select one, two and use all the sentences for each entity pair, respectively. Here we only select the top 100, 200, 300 precisions for our experiment. Experiment data of other methods (CNN+ATT, PCNN+ATT, Rank+ExATT) are obtained from their published papers. The results show our model outperforms all the baselines in P@N(One, Two, All, Average). Compared to Rank+ExATT, the latest state-of-the-state model, our model has a significant improvement on the average of P@100, P@200, P@300 by about 11.6\% on average.

\subsubsection{The effect of classifier number}

\begin{figure}[htbp]
\centering
\includegraphics[width=7cm]{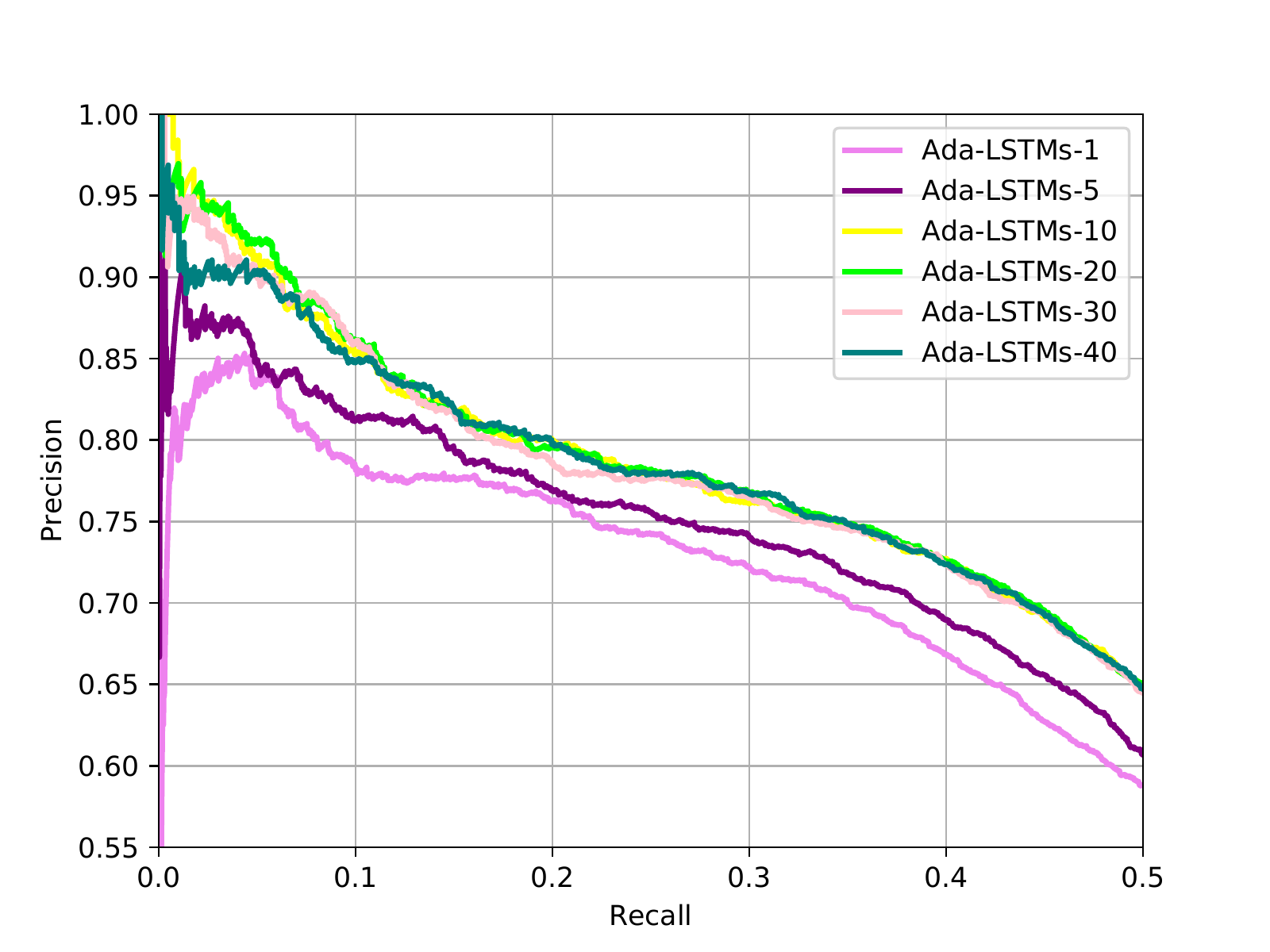}
\caption{Precision-Recall curves of different classifiers number.}
\label{fig:result_comp}
\end{figure}

To study the impact of the classifier number on our model performance, we set different numbers of classifiers. The result is given in Figure \ref{fig:result_comp}. One can see that when the classifier number of Ada-LSTMs is relatively small, the algorithm performance increases significantly with the increase of classifier number. Ada-LSTMs 10 $>$ Ada-LSTMs 5 $>$ Ada-LSTM 1. However, when the classifier number becomes large, the performance improvement gets less significant. The PR curves of Ada-LSTMs with 10, 20, 30, and 40 classifiers are quite similar. As mentioned, adaptive boosting plays two roles in our model due to ensembling the models and updating the gradient descent during the back propagation of neural networks via re-weighting.
%One can also see that our proposed model with adaptive boosting outperforms Att-LSTMs which does not apply adaptive boosting. 

%Thus, even though one Ada-LSTMs classifier is used in Ada-LSTMs, it still outperforms Att-LSTMs significantly, which 

\iffalse
%\begin{comment}
\begin{center}
\begin{tabular}{|c|c|c|c|c|c|c|}
\hline
\multicolumn{1}{|c|}{P@N(\%) }&
\multicolumn{1}{c|}{100}&
\multicolumn{1}{c|}{200}&
\multicolumn{1}{c|}{300}&
\multicolumn{1}{c|}{400}&
\multicolumn{1}{c|}{500}&
\multicolumn{1}{c|}{Avg}
\cr
\hline
\hline
Rank+ExATT 		&83.5 	&82.2 	&78.7 	&77.2 	&73.1 	&79.0\\
\hline
Ada+LSTM      	&\bf95.0 	&\bf92.5  	&\bf92.0 	&\bf91.7 	&\bf89.2 	&\bf92.0\\
\hline
\end{tabular}
\end{center}
%\end{comment}
\fi

\section{Conclusions}

In this paper, we proposed to integrate attention-based LSTMs with adaptive boosting model for relation extraction. 
Compared to the previous models, our proposed model is more effective and robust.
Experimental results on the widely used dataset show that our method significantly outperforms the baselines. In the future, it would be interesting to apply the proposed framework to other tasks, such as image retrieval and abstract extraction. 
%we will try more ensembles methods, such as random forest and bagging on relation extraction problem. Besides, 

\bibliographystyle{named}
\bibliography{ijcai18}

\begin{thebibliography}{}

\bibitem[\protect\citeauthoryear{De~Boer \bgroup \em et al.\egroup
  }{2005}]{de2005tutorial}
Pieter-Tjerk De~Boer, Dirk~P Kroese, Shie Mannor, and Reuven~Y Rubinstein.
\newblock A tutorial on the cross-entropy method.
\newblock {\em Annals of operations research}, 134(1):19--67, 2005.

\bibitem[\protect\citeauthoryear{Freund \bgroup \em et al.\egroup
  }{1996}]{freund1996experiments}
Yoav Freund, Robert~E Schapire, et~al.
\newblock Experiments with a new boosting algorithm.
\newblock In {\em Proceedings of The International Conference on Machine
  Learning}, pages 148--156, 1996.

\bibitem[\protect\citeauthoryear{Hoffmann \bgroup \em et al.\egroup
  }{2011}]{hoffmann2011knowledge}
Raphael Hoffmann, Congle Zhang, Xiao Ling, Luke Zettlemoyer, and Daniel~S Weld.
\newblock Knowledge-based weak supervision for information extraction of
  overlapping relations.
\newblock In {\em Proceedings of the 49th Annual Meeting of the Association for
  Computational Linguistics: Human Language Technologies}, pages 541--550,
  2011.

\bibitem[\protect\citeauthoryear{Kim and Kang}{2010}]{kim2010ensemble}
Myoung-Jong Kim and Dae-Ki Kang.
\newblock Ensemble with neural networks for bankruptcy prediction.
\newblock {\em Expert systems with applications}, 37(4):3373--3379, 2010.

\bibitem[\protect\citeauthoryear{Kingma and Ba}{2014}]{kingma2014adam}
Diederik Kingma and Jimmy Ba.
\newblock Adam: A method for stochastic optimization.
\newblock {\em arXiv preprint arXiv:1412.6980}, 2014.

\bibitem[\protect\citeauthoryear{Li \bgroup \em et al.\egroup
  }{2017}]{li2017cross}
Dandan Li, Shuzhen Yao, Senzhang Wang, and Ying Wang.
\newblock Cross-program design space exploration by ensemble transfer learning.
\newblock In {\em Proceedings of the 36th International Conference on
  Computer-Aided Design}, pages 201--208, 2017.

\bibitem[\protect\citeauthoryear{Lin \bgroup \em et al.\egroup
  }{2016}]{linneural}
Yankai Lin, Shiqi Shen, Zhiyuan Liu, Huanbo Luan, and Maosong Sun.
\newblock Neural relation extraction with selective attention over instances.
\newblock In {\em Proceedings of The 54th Annual Meeting of the Association for
  Computational Linguistics}, 2016.

\bibitem[\protect\citeauthoryear{Liu \bgroup \em et al.\egroup
  }{2016}]{liu2016learning}
Yang Liu, Chengjie Sun, Lei Lin, and Xiaolong Wang.
\newblock Learning natural language inference using bidirectional lstm model
  and inner-attention.
\newblock {\em arXiv preprint arXiv:1605.09090}, 2016.

\bibitem[\protect\citeauthoryear{Mintz \bgroup \em et al.\egroup
  }{2009}]{mintz2009distant}
Mike Mintz, Steven Bills, Rion Snow, and Dan Jurafsky.
\newblock Distant supervision for relation extraction without labeled data.
\newblock In {\em Proceedings of the Joint Conference of the 47th Annual
  Meeting of the ACL and the 4th International Joint Conference on Natural
  Language}, pages 1003--1011, 2009.

\bibitem[\protect\citeauthoryear{Miwa and Bansal}{2016}]{miwa2016end}
Makoto Miwa and Mohit Bansal.
\newblock End-to-end relation extraction using lstms on sequences and tree
  structures.
\newblock {\em arXiv preprint arXiv:1601.00770}, 2016.

\bibitem[\protect\citeauthoryear{Riedel \bgroup \em et al.\egroup
  }{2010}]{riedel2010modeling}
Sebastian Riedel, Limin Yao, and Andrew McCallum.
\newblock Modeling relations and their mentions without labeled text.
\newblock In {\em Proceedings of Joint European Conference on Machine Learning
  and Knowledge Discovery in Databases}, pages 148--163. Springer, 2010.

\bibitem[\protect\citeauthoryear{Rokach}{2010}]{rokach2010ensemble}
Lior Rokach.
\newblock Ensemble-based classifiers.
\newblock {\em Artificial Intelligence Review}, 33(1):1--39, 2010.

\bibitem[\protect\citeauthoryear{Surdeanu \bgroup \em et al.\egroup
  }{2012}]{surdeanu2012multi}
Mihai Surdeanu, Julie Tibshirani, Ramesh Nallapati, and Christopher~D Manning.
\newblock Multi-instance multi-label learning for relation extraction.
\newblock In {\em Proceedings of the 2012 Joint Conference on Empirical Methods
  in Natural Language Processing and Computational Natural Language Learning},
  pages 455--465, 2012.

\bibitem[\protect\citeauthoryear{Xu \bgroup \em et al.\egroup
  }{2015}]{xu2015show}
Kelvin Xu, Jimmy Ba, Ryan Kiros, Kyunghyun Cho, Aaron Courville, Ruslan
  Salakhudinov, Rich Zemel, and Yoshua Bengio.
\newblock Show, attend and tell: Neural image caption generation with visual
  attention.
\newblock In {\em Proceedings of The International Conference on Machine
  Learning}, pages 2048--2057, 2015.

\bibitem[\protect\citeauthoryear{Yang \bgroup \em et al.\egroup
  }{2016}]{yang2016hierarchical}
Zichao Yang, Diyi Yang, Chris Dyer, Xiaodong He, Alex Smola, and Eduard Hovy.
\newblock Hierarchical attention networks for document classification.
\newblock In {\em Proceedings of the 2016 Conference of the North American
  Chapter of the Association for Computational Linguistics: Human Language
  Technologies}, 2016.

\bibitem[\protect\citeauthoryear{Ye \bgroup \em et al.\egroup
  }{2017}]{ye2017jointly}
Hai Ye, Wenhan Chao, and Zhunchen Luo.
\newblock Jointly extracting relations with class ties via effective deep
  ranking.
\newblock {\em Proceedings of the 55th Annual Meeting of the Association for
  Computational Linguistics}, 2017.

\bibitem[\protect\citeauthoryear{Zeng \bgroup \em et al.\egroup
  }{2014}]{zeng2014relation}
Daojian Zeng, Kang Liu, Siwei Lai, Guangyou Zhou, Jun Zhao, et~al.
\newblock Relation classification via convolutional deep neural network.
\newblock In {\em Proceedings of 24th International Conference on Computational
  Linguistics}, pages 2335--2344, 2014.

\bibitem[\protect\citeauthoryear{Zeng \bgroup \em et al.\egroup
  }{2015}]{zeng2015distant}
Daojian Zeng, Kang Liu, Yubo Chen, and Jun Zhao.
\newblock Distant supervision for relation extraction via piecewise
  convolutional neural networks.
\newblock In {\em Proceedings of the 2015 Conference on Empirical Methods in
  Natural Language Processing}, pages 17--21, 2015.

\bibitem[\protect\citeauthoryear{Zhang \bgroup \em et al.\egroup
  }{2015}]{zhang2015bidirectional}
Shu Zhang, Dequan Zheng, Xinchen Hu, and Ming Yang.
\newblock Bidirectional long short-term memory networks for relation
  classification.
\newblock In {\em Proceedings of the 29th Pacific Asia Conference on Language,
  Information and Computation}, pages 73--78, 2015.

\bibitem[\protect\citeauthoryear{Zhang \bgroup \em et al.\egroup
  }{2017}]{zhang2017position}
Yuhao Zhang, Victor Zhong, Danqi Chen, Gabor Angeli, and Christopher~D Manning.
\newblock Position-aware attention and supervised data improve slot filling.
\newblock In {\em Proceedings of the 2017 Conference on Empirical Methods in
  Natural Language Processing}, pages 35--45, 2017.

\bibitem[\protect\citeauthoryear{Zheng \bgroup \em et al.\egroup
  }{2016}]{zheng2016aggregating}
Hao Zheng, Zhoujun Li, Senzhang Wang, Zhao Yan, and Jianshe Zhou.
\newblock Aggregating inter-sentence information to enhance relation
  extraction.
\newblock In {\em Proceedings of the 30th AAAI Conference on Artificial
  Intelligence}, pages 3108--3115, 2016.

\end{thebibliography}

\end{document}